\documentclass[prl,aps,showpacs,twocolumn]{revtex4}
\usepackage{dcolumn}
\usepackage{bm}
\usepackage{amsmath,amssymb,amsfonts,latexsym,fancyhdr,graphicx,epstopdf,times,txfonts}

\newcommand{\nbar}{\overline{n}}

\newcommand{\ket}[1]{\left\vert#1\right\rangle}

\newcommand{\pro}[3]{\left\vert#1\rangle_{#2}\langle#3\right\vert}

\newcommand{\bra}[1]{\left\langle#1\right\vert}

\begin{document}
\title{A simple trapped-ion architecture for high-fidelity Toffoli gates}
\author{Massimo Borrelli$^1$, Laura Mazzola$^{1,2}$, Mauro Paternostro$^2$, and Sabrina Maniscalco$^1$}
\affiliation{$^1$ Turku Centre for Quantum Physics, Department of Physics and Astronomy, University of
Turku, FI-20014 Turun yliopisto, Finland\\
$^2$ School of Mathematics and Physics, Queen's University, BT7 1NN Belfast, United Kingdom}

\begin{abstract}
We discuss a simple architecture for a quantum Toffoli gate implemented using 
three trapped ions. The gate, which in principle can be implemented with a single laser-induced operation, is effective under rather general conditions and is strikingly 
robust (within any experimentally realistic range of values) against dephasing, heating and random fluctuations of the Hamiltonian parameters. We provide a full characterization of the unitary and noise-affected gate using three-qubit quantum process tomography.
\end{abstract}
\pacs{03.67.Lx,32.80.Qk,37.10.Ty}

\maketitle

In the quest for scalability of a quantum computing device, the role played by many-qubit gates is quite central. {\it Adequate sets} of gates have been identified that allows for the break-down of complex computational networks in simpler tasks involving at most two qubits per time~\cite{barenco}. Unfortunately, the overhead in terms of the length of corresponding quantum circuits (i.e. the number of such elementary operations being required) soon overcomes the advantage provided by having to manage only two-body interactions. Multi-qubit gates exist, able to bypass such a problem by requiring the simultaneous conditional evolution of three or more qubits. Among these, the Toffoli gate~\cite{tommaso} is celebrated for its role in phase-estimation and error correction protocols~\cite{ec}, as well as in the quantum factorization algorithm~\cite{shor}. Remarkably, the Toffoli gate has jumped across the field of computing, from classical to quantum, playing an important role in schemes for reversible classical computation~\cite{tommaso2}. The recent effort put in the task of harnessing a three-qubit Toffoli gate has been considerable. On one side, important experimental demonstrations on the practical realization of such gate have been reported~\cite{monz,lanyon}. On the other hand, significant improvement in the design of economic ways of implementing an $n-$qubit Toffoli gate have arisen from realizing that less resources are needed when using particles that live in higher-dimenional Hilbert spaces~\cite{ralph,altri}. Interesting and compact architectures using the measurement-based paradigm for information processing have been suggested, in this context~\cite{pattame}.

In this paper, we present a protocol for the implementation of a three-qubit Toffoli gate in a trapped-ion architecture. Our scheme exploits an enlarged computational space consisting of three-level particles and an ancillary phononic mode~\cite{cinesi}. While information is encoded only in two electronic states of each ion, their third levels are used as convenient {\it working spaces}, similarly to the phononic ancilla. 
In this respect, our algorithm is close in spirit to the work by Ralph {\it et al.}~\cite{ralph}, although our protocol  is different by construction, and to further proposals exploiting higher-dimensional particles for improved manipulation of the computational states~\cite{altri}. In turn, such enlarged computational space allows a considerable reduction in the number of operations used in order to implement the three-qubit gate. In fact, our proposal requires roughly $44\%$ of the operations needed in the seminal experimental proof provided by Monz {\it et al.}~\cite{monz}. We thoroughly characterize the gate making use of three-qubit quantum process tomography (QPT)~\cite{NC} and reveal its striking robustness against leakage from the computational space, dephasing, heating and laser-power fluctuations. We thus provide an accurate platform for the experimental implementation of such a key gate in the design of scalable computing devices.

{\it The protocol.-}  Our system consists of $N{=}3$ ions in a linear electomagnetic trap.  We consider three internal energy levels $\{|l_j\rangle,|g_j\rangle,|e_j\rangle\}$ in a ladder configuration and the  centre-of-mass mode $a$ at frequency $\nu$. If the ions interact with an external classical laser at frequency $\omega_{L}$ it is possible to couple the external and internal degrees of freedom of the ions. In the interaction picture with respect to the energy of the internal degree of freedom and the vibrational mode, the Hamiltonian describing such coupling reads
\begin{equation}
\hat H_{I}(t){=}({\hbar\Omega}/{2})\hat\sigma_{-}^{(\alpha\beta)}e^{-i\eta(\hat ae^{-i\nu t}+\hat a^{\dagger}e^{i\nu t})-i(\omega_{\alpha\beta}-\omega_{L})t}{+}{h.c.},
\label{eq1}
\end{equation}
where $\Omega{\in}\mathbb{R}$ is the Rabi frequency of the transition $|\alpha\rangle{\leftrightarrow}|\beta\rangle$ (with $\alpha,\beta{=}e,g,l$), $\hat\sigma_{-}^{(\alpha\beta)}{=}|\alpha\rangle\langle\beta|$, $\omega_{\alpha\beta}$ is the corresponding transition frequency and $\eta$ is the Lamb-Dicke parameter~\cite{blattreview}. Moreover, we have introduced the phononic annihilation (creation) operator $\hat{a}~(\hat{a}^\dag)$ of the quantized phononic centre-of-mass mode. In the following, we will be exploiting the well-known flexibility in laser-induced trapped-ion dynamics that is achieved by tuning the laser-ion detuning  $\delta_{\alpha\beta}{=}\omega_{\alpha\beta}{-}\omega_{L}$~\cite{blattreview}. In the Lamb-Dicke regime defined by $\eta{\ll}1$, we can expand Eq.~\eqref{eq1} up to the second order in $\eta$ and set 
$\delta_{\alpha\beta}{=}\nu$ (i.e. we tune the laser to the first red sideband of the coupled spin-phonon system embodied by a single ion and the centre-of-mas mode). This engineers the energy-conserving coupling $\hat H^{(\alpha\beta)}_{r}(\zeta){=}\zeta\hat a\hat\sigma_{+}^{(\alpha\beta)}{+}h.c.$, where $\zeta{=}{\hbar\eta(\Omega}/{2})$ and a phononic excitation is created (destroyed)  upon annihilation (creation) of a spin quantum. Similarly, the choice $\delta_{\alpha\beta}{=}{-}\nu$ (corresponding to the tuning to the first blue sideband) induces the coupling 
$\hat H^{(\alpha\beta)}_{b}(\zeta){=}\zeta\hat a^{\dagger}\hat \sigma_{+}^{(\alpha\beta)}{+}h.c.$
where spin and phononic excitations are simultaneously created or destroyed. We now show how to realize a Toffoli gate using the unitary evolution given by the Hamiltonian $\hat{H}_r$, together with properly arranged single-qubit operations preceding and following the dynamics induced by $\hat H_r$. 
We will be working in the single-excitation sector of the Hilbert space of the whole ionic string, including the phononic mode. Such operations will be required to guarantee that the state of the system remains within such subspace. Moreover, we want to avoid the presence of correlations between the internal degrees of freedom of the string and their vibrational one. Therefore, starting from a phononic mode with no excitations, we have to enforce that when the Toffoli gate is completed, the centre-of-mass mode is back to its vacuum state.

We are now in a position to describe the details of our protocol. First, we codify three qubits in the internal degrees of freedom of the ions, by using the simple encoding scheme $(\ket{0_1},\ket{1_1}){=}(\ket{g_1},\ket{e_1})$, $(|{0_j}\rangle,|{1_j}\rangle){=}(|{g_j}\rangle,|{l_j}\rangle)~(j{=}2,3)$.
With this, we construct an eight-state basis for the three-qubit system as
${\cal B}{=}\{|000\rangle,|100\rangle,|010\rangle,|110\rangle,|001\rangle,|101\rangle,-i|011\rangle,-i|111\rangle\}_{123}$, where we have redefined the last two states so as to include an overall phase factor (the choice is made only to simplify our calculations).
Our protocol begins with the realization of 
the single-qubit operations on the qubit $1$-phononic mode system given by ($\tau{=}\pi/2\zeta$)
\begin{equation}
\label{hac}
\begin{aligned}
\hat R_{A}^{+}({\pi}/{2\zeta}){=}e^{i\hat H^{(eg)}_b(\zeta)\tau},
\hat R_{B[C]}^{-}({\pi}/{2\zeta}){=}e^{i\hat H^{(le[g])}_r(\zeta)\tau},
\end{aligned}
\end{equation}
The first transformation in Eq.~(\ref{hac}) excites the first blue sideband for the $|g_1\rangle{\leftrightarrow}|e_1\rangle$ transition while the remaining two embody the first red sideband excitations for the $|e_1\rangle{\leftrightarrow}|l_1\rangle$ and $|g_1\rangle{\leftrightarrow}|l_1\rangle$ passages. In order to work with the same phononic mode, we need laser fields with frequencies $\omega_{L}^{A}{-}\omega_{ge}{=}{-}\nu, \omega_{L}^{B}{-}\omega_{le}{=}\nu, \omega_{L}^{C}{-}\omega_{gl}{=}\nu$. In the subspace with at most a single excitation, it is straightforward to see that the composite operation $\hat{\cal R}{=}\hat R_{C}^{-}\left({\pi}/{2\zeta}\right)\hat R_{B}^{-}\left({\pi}/{2\zeta}\right)\hat R_{A}^{+}\left({\pi}/{2\zeta}\right)$, operated on states having initially no phononic excitations, performs a logical $\hat\sigma_x$ gate in the space of the phononic mode, controlled by the spin state $\ket{g_1}$. That is
\begin{equation}
\label{eq4}
\hat{\cal R}
\begin{pmatrix}
\ket{g_1,0}\\
\ket{e_1,0}
\end{pmatrix}{=}
\begin{pmatrix}
\ket{g_1,1}\\
\ket{e_1,0}
\end{pmatrix}.
\end{equation}
It is important to notice that this operation encodes a logical qubit in the single-excitation pair of states $\{|g_1,1\rangle,|e_1,0\rangle\}$. Therefore, at this stage, one has to consider the attainable states of the extended system comprising both the internal and external degree of freedom of the ion-string and having a single overall excitation. This is re-interpreted, in our scheme, as the computational space of three logical qubits, one of which being embodied, in a sort of {\it dual rail} encoding, by $\{|0_L\rangle,|1_L\rangle\}{\equiv}\{|g_1,1\rangle,|e_1,0\rangle\}$~\cite{NC}.
We now couple each ion to a field of frequency $\omega_{L}{=}\omega_{eg}{+}\nu$ at Rabi frequency $\Omega_{j}$. The total interaction is described by the Tavis-Cummings model~\cite{TC} 
\begin{equation}
\hat H_{\textrm{TC}}=\sum_{j=1}^{3}({\hbar\eta\Omega_{j}}/{2})\hat a|e_{j}\rangle\langle g_{j}|+h.c.
\label{eq6} 
\end{equation}
We let each of the single-excitation states mentioned above evolve under $\hat{H}_{\text{TC}}$ for a time $t_T$ at which we turn Eq.~\eqref{eq6} off and apply the gates forming $\hat{\cal R}$ in reverse order. This separates the state of the centre-of-mass mode from the spin state of ion $1$. That is, we decode the logical qubit $\{|0_L\rangle,|1_L\rangle\}$ resetting the vibrational mode into $\ket{0}$ and remaining with just the ion-string computational states. The vibrational degrees of freedom can now be traced out without affecting, in principle, the resulting three-qubit gate   
\begin{equation}
\hat{\cal U}_{T}(t)=(\hat {\cal R}^{\dagger}\otimes\hat\openone_{23}){\exp}[-({i}/{\hbar})\hat H_{\rm{TC}}t_{T}](\hat{\cal R}\otimes\hat\openone_{23}),
\label{t}
\end{equation}
where $\hat\openone_{23}$ is the identity operators in the tensor-product space of qubits $2$ and $3$. The main idea now is finding an instant of time $t_{T}$ such that $\hat{\cal U}_{T}(t)$ is as close as possible to the Toffoli gate $\hat{T}{=}\hat{\openone}_1{\otimes}(\hat{\openone}_{23}{-}\ket{11}_{23}\!\bra{11}){+}\hat\sigma_{x,1}{\otimes}\pro{11}{23}{11}$.
We first observe that there are four time-scales associated with the dynamics at hand. Each of them is determined by the inverse of the Rabi frequencies of the processes involved in this scheme, i.e. $\Theta_{123}{=}\hbar\eta(\sum^3_{j=1}\Omega^2_j)^{1/2}
, \Theta_{1j}{=}\hbar\eta(\Omega_{1}^{2}{+}\Omega_{j}^{2})^{1/2}~(j{=}2,3)$ and $\Theta_{1}{=}\hbar\eta\Omega_{1}$. The pedices used in such expressions identify the qubits that participate to the interaction with the phononic mode.  
It is thus clear that the quest for $t_{T}$ is equivalent to the research of a set of suitable single-ion Rabi frequencies $\{\Omega_{j}\}$ such that our goal is achieved. An extensive numerical optimization (performed in the spirit of optimal control theory) leads to the following relative ratios of coupling strengths 
$\Omega_{1}{:}\Omega_{2}{:}\Omega_{3}{=}1{:}\sqrt{143}{:}16$. With this at hand, 
at the optimal instant of time given by $t_T{=}\pi/\eta\Omega_{1}$, $\hat{\cal U}_{T}$ take the following matrix form [in the ordered computational basis $\{\ket{i}\}~(i{=}1,..,8)$. Here $i$ stands for the decimal-number value of each element of the three-ion basis ${\cal B}$]
\begin{equation}
\hat{\cal U}_{T}{=}
\begin{pmatrix}
  1 & 0 & 0 & 0 & 0 & 0 & 0 & 0 \\
  0 & 1 & 0 & 0 & 0 & 0 & 0 & 0 \\
  0 & 0 & 1 & 0 & 0 & 0 & 0 & 0 \\
  0 & 0 & 0 & 1 & 0 & 0 & 0 & 0 \\
  0 & 0 & 0 & 0 & 0.999 & 0 & 0 & 0 \\
  0 & 0 & 0 & 0 & 0 & 0.998 & 0 & 0 \\
  0 & 0 & 0 & 0 & 0 & 0 & 0 & 1 \\   
  0 & 0 & 0 & 0 & 0 & 0 & 1 & 0 \\
\end{pmatrix}.
\label{tmatrix}
\end{equation}
Needless to say, other choices can be found for the set of Rabi frequencies that achieve a gate close to a Toffoli one. However the latter cannot be exactly achieved due to the necessity of maximize, simultaneously, eight trigonometric functions of incommensurate frequencies. Clearly, the only important parameter in our model is the ratio of the Rabi frequencies rather than their actual value. 
The time needed in order to implement the whole gate is 
$t_{G}=({\pi}/{\eta})[2\!\sum_{k{=}a,b,c}\!\Omega^{-1}_k{+}\Omega^{-1}_1]$,
where $\Omega_{k}$ is the Rabi frequency of pulse $k{=}a,b,c$ in Eq.~\eqref{hac}. 
It is crucial to stress that in such unitary picture, the implementation of Eq.~\eqref{eq6} for a time $t_T$ would be sufficient to implement a full Toffoli gate over the logical target qubit $\{|0_L\rangle,|1_L\rangle\}$, which shows the striking economic nature of our proposal. Clearly, the use of excited vibrational states would open the protocol to the effects of phononic heating and losses. The necessity of removing such excitation (so as to make the gate robust) motivates the use of the encoding-decoding steps given by $\hat{\cal R}$.
\begin{figure}[b]
\includegraphics[width=1.0\linewidth]{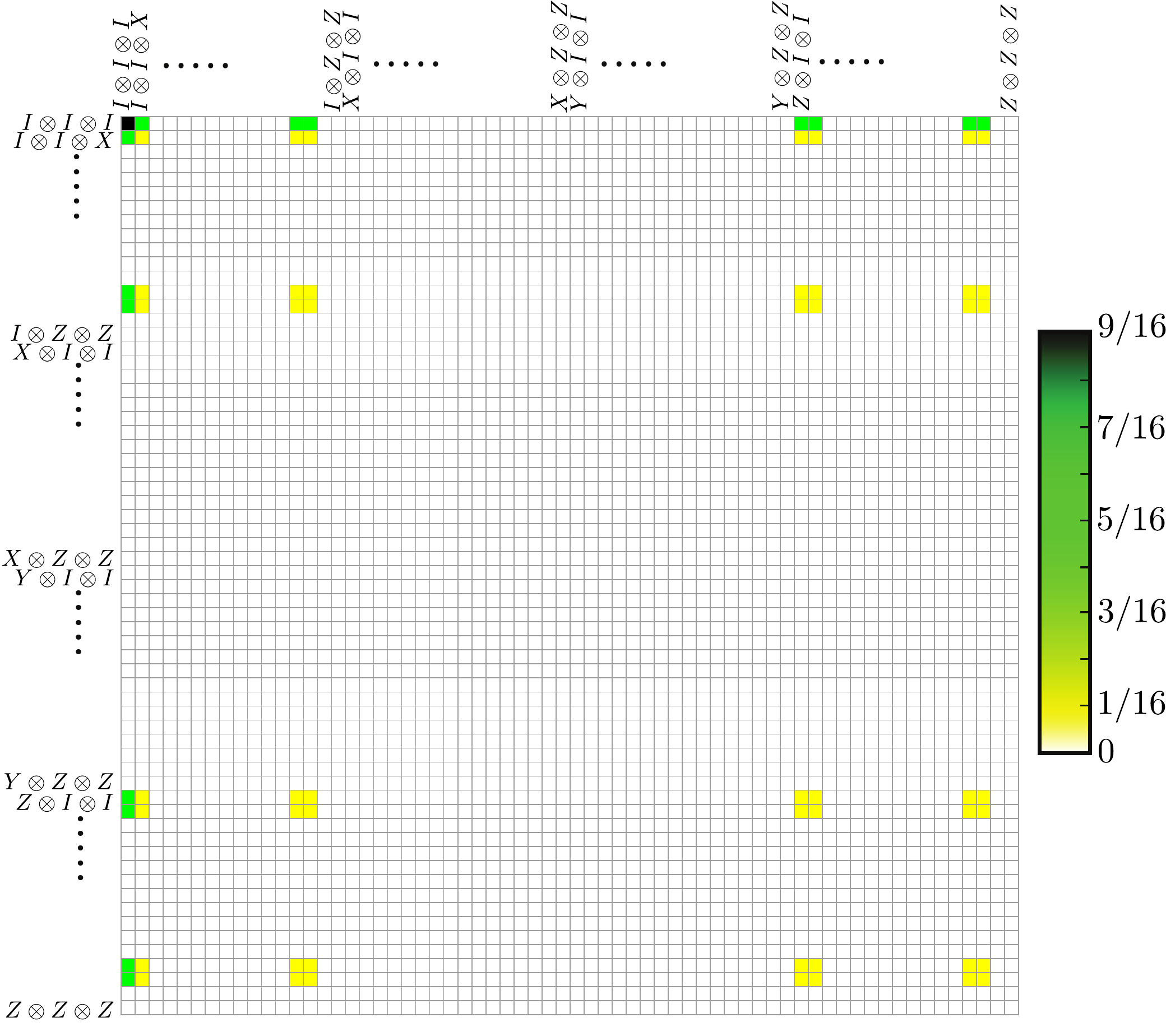}
\caption{(Color online) Reconstructed process matrix corresponding to the unitary process $\hat{\cal U}_T$. The matrix is expressed in the three-qubit operator basis formed by the elements $\{I{\equiv}\hat\openone,X{\equiv}\hat\sigma_x,Y{\equiv}-i\hat\sigma_y,Z{\equiv}\hat\sigma_z\}$. We show the moduli of the matrix entries. The differences with respect to the elements of an ideal gate are ${\cal O}(10^{-4})$.}
\label{Pmat}
\end{figure}

To evaluate the quality of our proposal, the dynamics encompassed by the physical processes described so far should be characterized in a state-independent way. In what follows, we take an experiment-inspired approach and use QPT~\cite{NC} as the tool to estimate the performance of the gate. Any completely positive $N$-qubit map $\Phi(t)$ is specified by a set of $4^N$ orthogonal operators $\{\hat{\cal K}_m\}$ such that 
\begin{equation}
\label{determino}
{\Phi}(t)\varrho(0){=}\sum_{m,n}\chi_{mn}(t)\hat{\cal K}_m\varrho(0)\hat{\cal K}^{\dag}_n,
\end{equation}
where $\rho(0)$ is the initial density matrix of the system and we have introduced the $4^N{\times}4^N$ time-dependent {\it process matrix} ${\bm \chi}(t)$ that incorporates full information on the details of the evolution embodied by $\Phi(t)$. Pragmatically, this description is of the utmost usefulness as the map is fully specified by considering only a fixed set of operators, whose knowledge determines the process matrix. The knowledge of the process matrix allows to evaluate the closeness of the mechanism under scrutiny to the ideal one corresponding to $\hat T$ (with process matrix ${\bm\chi}_T$) by means of the {\it gate fidelity} ${\cal F}_g(t_G){=}{\rm Tr}[{\bm \chi}_{T}{\bm \chi}(t_G)]$. In turn, this is useful to determine the average state fidelity associated with gate $\overline{\cal F}_s(t_G)$. This provides a state-independent indicator of the quality of the gate by averaging the fidelity between the ideal and actual output states over all pure inputs. For a $d$-dimensional problem ($d{=}2^N$) we have~\cite{JLO}
$\overline{\cal F}_s(t_G){=}(d{\cal F}_g(t_G){+}1)/(d{+}1)$.
In Fig.~\ref{Pmat} we show the representation of the reconstructed process matrix in the tensorial operator-basis constructed by considering the single-qubit operators $\{\hat\openone,\hat\sigma_x,-i\hat\sigma_y,\hat\sigma_z\}$. The entries of ${\bm \chi}(t_G)$ differ from those of the ideal one by ${\cal O}(10^{-4})$, thus witnessing the excellent quality of the gate we have achieved. An additional confirmation comes from achieving an average {\it infidelity} $1{-}\overline{\cal F}_s(t_G){\sim}10^{-5}$.

{\it Analysis of imperfections.-} So far, we have considered only unitary evolutions. In order to provide an estimate of the efficiency of the gate under realistic experimental conditions, 
we should consider the most severe sources of imperfections in the ion-trap architecture addressed here~\cite{monz}. In the following, we concentrate on quality-limiting effects of a non-technical nature and take into account decoherence of the quantum information stored in the phononic mode given by the vibrational centre-of-mass mode of the ion-string as well as heating due to the coupling between the phononic mode and a bath at finite temperature. The possibility to tune the Rabi frequency and the semiclassical approach used in the ion-pulse dynamics (implying the use of intense fields in the construction of $\hat{\cal R}$) allow us to neglect the time needed to implement the single-qubit operations Eq.~\eqref{hac}. We thus consider the master equation 
\begin{equation}
\begin{aligned}
&\partial_t{\rho}(t){=}{-}i[\hat H_{\text{TC}},\rho(t)]{-}(\kappa/2)(\bar{n}{+}1)(\hat a^{\dagger}\hat a\rho(t){+}\rho(t) \hat a^{\dagger}\hat a{-}2\hat a\rho(t) \hat a^{\dagger})\\
&-({\kappa\bar{n}}/{2})(\hat a\hat a^{\dagger}\rho(t)+\rho(t)\hat a\hat a^{\dagger}-2\hat a^{\dagger}\rho \hat a){-}\gamma[\hat a^{\dagger}\hat a,[\hat a^{\dagger}\hat a,\rho(t)]]
\label{me1}
\end{aligned}
\end{equation}
where $\rho(t)$ is the density matrix of the ionic string including the centre of mass mode, $\kappa$ is the heating rate, $\nbar$ is the mean number of phononic quanta of the bath at a given temperature and $\gamma$ is the dephasing rate. Analogously to the unitary case, the dynamical map ${\cal E}_{H}$ arising from Eq.~\eqref{me1} should be preceded and followed by the $\hat{\cal R}$ gate. That is, any initial state $\rho(0)$ of the three-ion system 
evolves until time $t_G$ according to 
\begin{equation}
\rho(t_{G}){=}\hat {\cal R}^{\dagger}[\mathcal{E}_{{H}}(\hat {\cal R}\rho({0})\hat {\cal R}^{\dagger})]\hat {\cal R}.
\label{rhot}
\end{equation}
The resulting open-system dynamics implies, in principle, leakage from the computational space and thus the spoiling of the desired Toffoli gate. In particular, the thermal evolution included in our assessment could lead us to abandon the subspace where the bosonic mode is in its vacuum state. 
The effective occurrence of such events and their influences over the average performance of our operations can be estimated by determining again the closeness of the map in Eq.~\eqref{rhot} to the ideal transformation $\hat T$. 
We have thus used our toolbox for QPT to quantify the average gate fidelity for the noise-affected evolution and estimate the deviations from ideality.

\begin{figure}[t]
\includegraphics[width=0.7\linewidth]{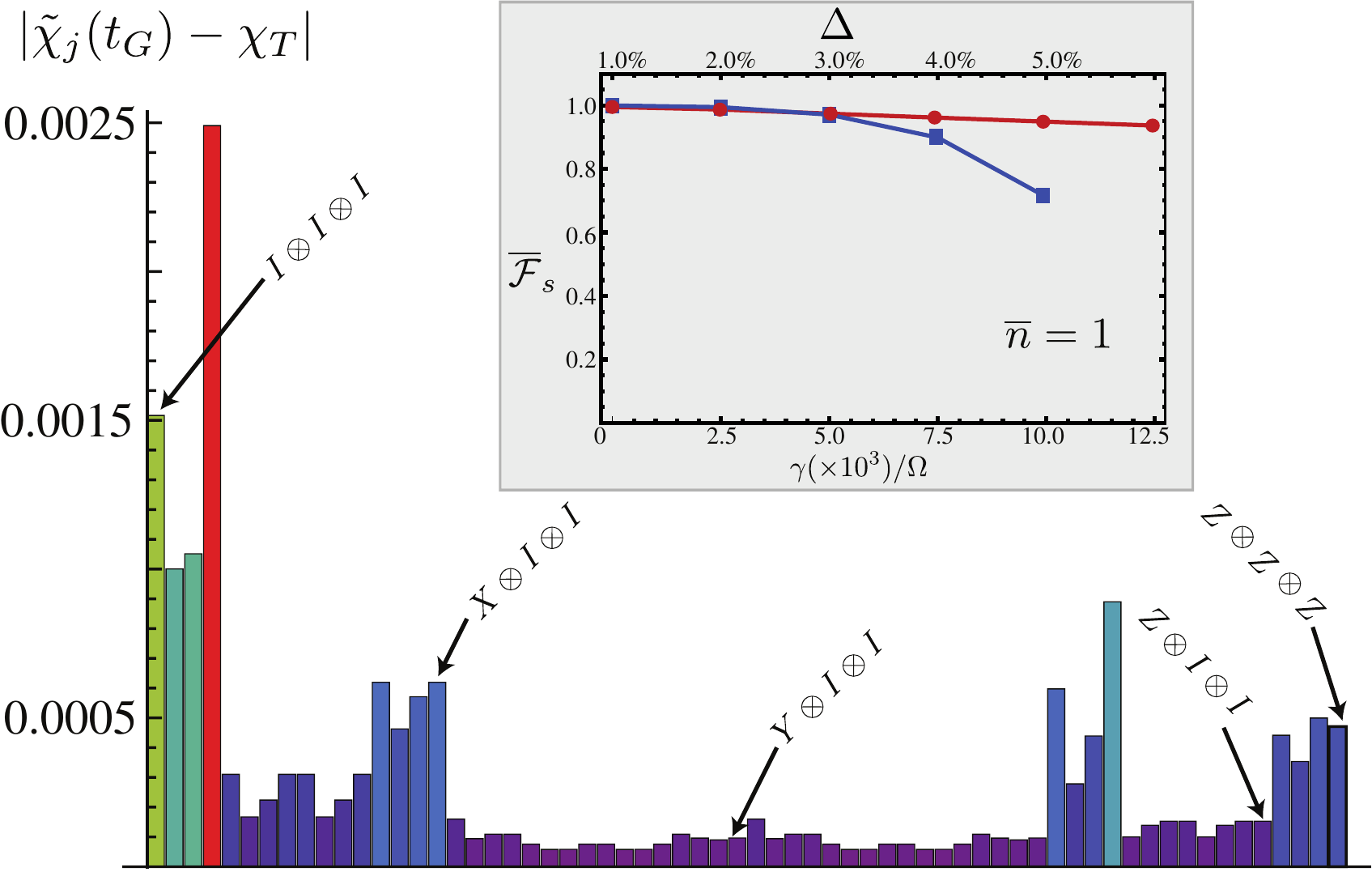}
\caption{(Color online) Main panel.- We take the largest entry per row in the discrepancy matrix $|\tilde{\bm \chi}(t_G){-}{\bm \chi}_T|$ for $\nbar{=}5,\,\gamma/\Omega{=}10^{-3}$. We have highlighted the bars corresponding to some of the operator-basis elements. Inset.- Lower horizontal axis and circle-shaped points: Average state fidelity for the $\hat{\cal U}_T$ gate against the dephasing $\gamma/\Omega$. At $\gamma{=}0$ it is $\overline{\cal F}_s{=}0.999988$, while for the larger dephasing rate that we have considered we have $\overline{\cal F}_s{>}0.93$. Upper horizontal axis and square-shaped points: Average gate fidelity for $\hat{\cal U}_T$ against the variance $\Delta$ of the uniform distribution determining the amplitudes of laser fluctuations in our model. The solid lines are only guides to the eye.}
\label{quant}
\end{figure}

We have first considered the effects of heating of the phononic mode due to noisy electric potentials on the surface of the trap electrodes, resulting in an effective bath at non-zero temperature. We have taken $(\kappa^{-1},\gamma^{-1}){=}(140,85)$ms and considered increasing values of $\nbar$, taking as reasonable maximum number of thermal excitations equal to 5~\cite{monz}. In such a {\it worst case scenario}, we have reconstructed the process matrix $\tilde{\bm \chi}(t_G)$ and checked its resemblance to ${\bm \chi}_T$. We have calculated the discrepancy $|\tilde{\bm \chi}(t_G){-}{\bm\chi}_T|$ and taken its maximum value per row of this matrix (see the main panel of Fig.~\ref{quant}). The largest deviation out of the $64$ values gathered in this way is a rather small $\simeq{2.5}{\times}10^{-3}$. In fact, the evaluation of the average gate fidelity leads to $\overline{F}_{s}{=}0.994855$, which is $99.5\%$ the value achieved for $\nbar{=}1$. The remarkable insensitivity of the scheme to the effects of an increased mean phonon number is therefore fully proven. Our analysis thus leads us to conclude that map ${\cal E}_H$ results in an effective dynamics that is approximated in an excellent way by 
$\rho(t_{G})\approx\rho_{q}(t_{G})\otimes|0\rangle\langle0|$, 
where $\rho_{q}(t_{G})$ is the density matrix of the three-ion system. 

Our next step is the evaluation of the dephasing effects, which has been performed by solving the master equation~(\ref{me1}) for $\nbar{=}1$, $\kappa^{-1}{=}140$ms and for increasing values of $\gamma$. The inset of Fig.~\ref{quant} (circle-shaped points) shows the quasi-independence of the effective gate from an increase in $\gamma$ by almost one order of magnitude with respect to the value estimated in recent cutting-edge experiments ($1{-}\overline{\cal F}_s{\in}[10^{-5},0.07]$ for $\gamma/\Omega\in[0,12.5]{\times}{10^{-3}}$)~\cite{monz}.

As remarked above, a key point in our proposal is the maintenance of precise ratios of the Rabi frequencies of the operations involved in the construction of $\hat{\cal U}_T$. Lasers fluctuations may, in principle, jeopardize the stability required in the scheme we have proposed. Although, experimentally, such contributions account for only a negligible percentage of the overall gate fidelity~\cite{monz}, the central role played by the Rabi frequency ratios in our protocol have prompted us to investigate them thoroughly. 
We have thus solved Eq.~\eqref{me1} again, this time treating the $\Omega_{j}$'s as stochastic variables which randomly oscillate around the corresponding ideal values. More in details, we have taken  $\Omega_{j}^{'}{=}\Omega_{j}{+}\delta\Omega_{j}$, with $\delta\Omega_j/\Omega_j$ a uniformly-distributed zero-mean variable with variance $\Delta{\in}[1,5]\%$. Using a sample of $500$ randomly drawn values of $\delta\Omega_j$ and evaluating the corresponding dynamical evolution, we have calculated the sample-averaged $\overline{\cal F}_s$. In the worst case scenario given by $\Delta{=}5\%$ (which grossly overestimates the current experimental capabilities), we have achieved an average fidelity of ${\simeq}71\%$ (see the square-shaped points in the inset of Fig.~\ref{quant}).

{\it Conclusions.-} We have discussed a scheme for the implementation of a three-qubit Toffoli gate that requires only $44\%$ of the total number of operations needed by a very recent experimental demonstration in a trapped-ion system. Despite using working space that lies outside the Hilbert space of three qubits, our protocol is remarkably robust and affected by only negligible leakage from the computational space. Moreover, it enjoys the gate-catalysis effects provided by the use of higher-dimensional information carriers. We have characterized the performances of the gate from an experimentally oriented viewpoint and making use of powerful tools typical of QPT. On the one hand, this has allowed an agile quantification of the overall gate quality. On the other hand, we have rigorously estimated the influences of the most relevant causes of gate imperfections in the setup at hand, finding quite a striking resilience. Economic schemes such as ours will be important in the design of experimental-friendly architectures for trapped-ion quantum computing that can find a rather non-demanding implementation in state-of-the-art settings~\cite{monz,monroe}. 

{\it Acknowledgements.-} MB, LM, and SM acknowledge  financial support from the Emil Aaltonen Foundation, the Finnish Cultural Foundation, the Magnus Ehrnrooth Foundation and the Turku Collegium of Science and Medicine. MP is grateful to the Centre for Quantum Physics, University of Turku, for the kind hospitality and acknowledges financial support from EPSRC (EP/G004579/1) and the British Council.

\end{document}